\documentclass[twocolumn,preprintnumbers,amsmath,amssymb,floatfix]{revtex4}

\usepackage{amssymb,amsmath}
\usepackage{graphicx}
\usepackage{dcolumn}
\usepackage{bm}

\def\A{{\cal{A}}}

\def\bra#1{ \langle{#1}| }

\def\opket#1#2 {  {#1} |{#2}\rangle }
\def\braopket#1#2#3{ \langle{#1}| {#2} |{#3}\rangle }

\def\d{{\rm d}}

\def\det{{\rm det}\,}

\def\e{{\rm e}}

\def\F{{\cal F}}

\def\ha{{1 \over 2}}

\def\H{{\cal H}}
\def\hh{\hat{h}}

\def\in{{\rm in}}

\def\K2{{\cal K}}
\def\ket#1{ |{#1}\rangle }

\def\p{{\bf p}}

\def\q{{\bf q}}

\def\R{{\hat{\cal R}}}

\def\Re{{\rm Re}}

\def\T{{\hat{\cal T}}}

\def\Wt{{\cal W}}

\def\spinor#1#2{\left( \begin{array}{c} {#1}\\
			                {#2} \end{array} \right)}

\def\zetab{\mbox{\boldmath$\zeta$\unboldmath}}
\def\zetabs{\mbox{\boldmath$\scriptstyle\zeta$\unboldmath}}

\def\sbar{\big/}
\def\dbar{ {\,\sbar\!\sbar} }
\def\dbarr{ {\,\sbar} }

\def\opsket#1#2#3 {  {#1} \dbar_{#2}{#3}\rangle }

\def\opssket#1#2#3 {  {#1} \dbarr_{#2}{#3}\rangle }

\newcommand{\sech}{\mathop{\rm sech}\nolimits} %

\begin{document}
\title{Uniform approximation of barrier penetration in phase space}
\author{Christopher S. Drew}
\affiliation{School of Mathematical Sciences,
University of Nottingham, University Park, Nottingham, NG7~2RD, UK}
\author{Stephen C. Creagh}
\email{stephen.creagh@nottingham.ac.uk}
\affiliation{School of Mathematical Sciences,
University of Nottingham, University Park, Nottingham, NG7~2RD, UK}
\author{Richard H. Tew}
\affiliation{School of Mathematical Sciences,
University of Nottingham, University Park, Nottingham, NG7~2RD, UK}
\date{Draft version \today}
\smallskip
\begin{abstract} A method to approximate transmission probabilities
for a nonseparable multidimensional barrier is applied to a waveguide
model. The method uses complex barrier-crossing orbits to represent reaction
probabilities in phase space and is uniform in the sense that it applies at
and above a threshold energy at which classical reaction switches
on. Above this threshold the geometry of the classically reacting
region of phase space is clearly reflected in the quantum
representation. Two versions of the approximation are applied.
A harmonic version which uses dynamics linearised around an instanton
orbit is valid only near threshold but is easy to use. A more accurate
and more widely applicable version using nonlinear dynamics is also
described.
\end{abstract}
\maketitle
\section{Introduction}
Semiclassical approaches to multidimensional tunnelling lead to very
interesting problems in complexified classical dynamics, often with
incompletely understood solutions. For example, recent work in references
\cite{Shudo,aY00,Takahashi,TYK,Onishi,Julia} has shown that nontrivial
geometrical structure such as complex homoclinic intersections have an
important role to play in multidimensional barrier penetration and that
even complex chaos can be relevant. Given the difficulty inherent in a
systematic treatment of multidimensional tunnelling as a result of such
issues,  it is perhaps surprising that a relatively simple description
can be given of barrier penetration at a critical energy where classically
allowed transmission mechanisms turn on and where primitive semiclassical
approximations must be replaced by somewhat more complicated
uniform ones.

An approach which achieves this has been proposed in references
\cite{sC04} and \cite{sc05} and in this paper we apply the
method explicitly to a model waveguide problem. The model
is chosen to be rather simple so that fully quantum calculations
are easy to perform accurately for purposes of comparison. We emphasise,
however, that semiclassical aspects of the calculation are as easily
applied to other problems, provided the topology is similar, and provide
a description, for example, of collinear atom-diatom reactions.
For that reason we use the terminology of chemical reactions
in this paper and equate the probability of transmission with a
probability of reaction. In fact the approach we describe here
provides a natural means of visualising the quantum scattering
problem in phase space and as such shows an interesting connection with
classical transition state theories of chemical reaction. These classical
theories have recently been of interest because the periodic-orbit dividing
surface (PODS) construction \cite{Pechukas, PODS2} has been generalised to
arbitrary dimensions using the construction of normally hyperbolic
invariant manifolds (NHIMS) \cite{Jaffe,Uzer,WW,HCN,LW}. The classical
constructions emerge  naturally in our
semiclassical approximation and we note that even though the
illustrations offered here are in two degrees of freedom there
are straightforward generalisations to higher dimensional problems
where the full generality of the NHIM construction comes into play.

The approximation we use can be stated very simply
as an abstract operator equation but for explicit illustration
we present results in phase space, using the Wigner-Weyl calculus.
 In particular,
we define a {\it Weyl symbol} of a transmission matrix
which represents, in an averaged sense, a reaction probability as a
function of phase space.
We find that above threshold the support of this Weyl symbol
closely mimics the shape of the classically reacting region
but the Weyl symbol itself also incorporates tunnelling and other
quantum effects. Notation and details for this construction are set
out in section~\ref{secps}.

 We implement two versions of the
theory. First, a harmonic approximation derived in \cite{sC04} is
applied in section~\ref{harmonic} whose classical input consists
simply of an instanton
orbit, along with its action  and monodromy matrix. This approximation
works when the classically reacting region is a small neighbourhood
of the initial condition for the instanton orbit and is harmonic
in the sense that it uses linearised dynamics generated by an elliptic
quadratic Hamiltonian on the Poincar\'e section. This harmonic version
covers the threshold case where classical reaction switches on as a
function of energy and is relatively easy to apply. It fails however
when the energy is too far above threshold and the classically reacting
region is too large to be adequately described by linearised dynamics.

A semiclassical approximation that is more accurate and has greater
range has been derived in \cite{sc05} and this is applied in
section~\ref{nonlinear}. This version uses fully nonlinear dynamics
to extend further from the instanton orbit. Like the harmonic version
it can be stated quite simply as an abstract operator equation but its
practical implementation is more difficult. Difficulty arises
primarily because we must invert an operator constructed
semiclassically as an evolution operator and this inversion cannot
at present be achieved in closed form. In this paper we achieve that
inversion using numerical methods and while this aspect of the
approach needs further work to provide an appealing semiclassical
method, we can verify unambiguously that the nonlinear version of the theory
is capable of describing the quantum transmission problem very
accurately (see Figure~\ref{figconj1}).

We conclude this section by outlining how these results
relate to existing work. The basic formalism here of relating
scattering to complex barrier-crossing orbits goes back to the
work of Miller and coworkers \cite{MG,MillerSmatrix}
on the classical $S$-matrix. Our
intent is to describe a simple uniform extension of
this approach which applies at the boundary of classical
reaction where the fate of classical orbits changes discontinuously.
What allows us to make progress is that we do not directly describe
the $S$-matrix but instead consider a transmission matrix derived
from it which gives probabilities rather than amplitides. The
advantage of this problem is that contributing orbits at the
boundary of classical reaction depend smoothly on initial
conditions, despite the singular nature of real orbits there
\cite{sC04}, and give simple semiclassical expressions.
This uniformisation is similar to established results relating one-dimensional
transmission probabilities  \cite{FW,BM,Smil} or the cumulative
reaction probability \cite{Millermicro} to sums over multiple barrier
crossings but includes information about how the probabilities depend
on the incoming state.
It is different however from uniform approximations
of the scattering operator such as described in \cite{ElranKay}
which describe explicit matrix elements. These are uniform with respect
to variation of quantum numbers whereas our approach treats the
scattering operator abstractly and is uniform with respect to energy
and in phase space.

Direct approximation of the scattering problem by
complex trajectories has recently been examined in \cite{aY00,Takahashi,TYK}
in the context of nonintegrable systems. It has been found there
that, while intuitively one might expect tunnelling processes to be
dominated by short complex orbits which cross the barrier directly,
the dominant complex orbits can have have a surprisingly nontrivial
topology  in the deep tunnelling regime. It has even been found
that the dominant complex orbits may be chaotic
in related treatments of quantum propagation \cite{Shudo,Onishi,Julia}.
We also find evidence of the ``fringed tunnelling'' characteristic
of such mechanisms in our fully quantum solutions but the theory
we outline is intended to cover only the immediate vicinity of
the reacting region
where direct tunnelling mechanisms are dominant. In the deep tunnelling
regime our uniform results revert to standard primitive approximations
and we should in principle be able to marry our approach with that of
\cite{aY00,Takahashi,TYK}. It is not obvious, however, that
a fully uniform calculation could easily be applied when
the contributing complex orbits are more numerous and more
complicated and we do not consider that problem explicitly
here.

\section{Representing the scattering matrix in phase space}\label{secps}
In the following sections we will develop semiclassical approximations
for representations of the scattering matrix in phase space. In this
section we illustrate these representations numerically using a
two-dimensional waveguide, which serves as a simplified model of a
collinear atom-diatom collision. This model consists of
a particle of unit mass moving in the potential
\begin{equation}
  \label{toypot}
  V(x,y)=\sech^2{x}+\frac{1}{2}\omega^2(x)(y-a(x))^2,
\end{equation}
where
\begin{eqnarray*}
  \omega^2(x)&=&\Omega^2+\lambda \sech^2{x}\\
  a(x)&=&\mu \sech^2{x}.
\end{eqnarray*}
A very similar potential has been used in \cite{aY00,TYK}.
The simplicity of this model will enable us easily to obtain
accurate numerical solutions, which will be useful for later comparison
with semiclassical approximations where exponentially small tunnelling
effects are of interest. We emphasise, however, that none of the
theory that follows is dependent on this simplicity and semiclassical
aspects of the discussion can just as easily be applied to any other
system, as long as the Hamiltonian is an analytic function of its
arguments. The essential structural features we assume are that the waveguide
should have a single bottleneck separating asymptotically decoupled
channels, which we label as reactant and product channels
respectively, and that the energy should be sufficiently close to threshold
that recrossings of the transition state do not occur.

\begin{figure}
  \begin{center}
	\includegraphics[width=7.5cm]{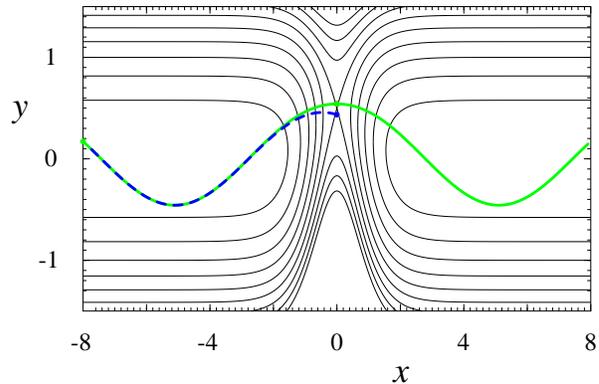}
  \end{center}
  \caption{\label{fig:orbit}Contours of the model potential are shown
for the case $\Omega=1$ and $\lambda=\mu=1/2$. The continuous line
shows a real trajectory which is an extension of the complex periodic
orbit (or instanton) used in the next section to construct the simplest
semiclassical approximation. The dashed curve shows the trajectory
defined by the decoupled dynamics of $V_\infty(y)$ to which it asymptotes.}
\end{figure}

For future reference, it will be useful
to denote the asymptotically decoupled potential by the symbol
\[
  V(x,y)\sim V_\infty(y)=\frac{1}{2}\Omega^2 y^2.
\]
We can therefore write the asymptotic scattering states for this
problem analytically, as solutions of a harmonic oscillator.

\begin{figure*}[ht]
\includegraphics[width=18cm]{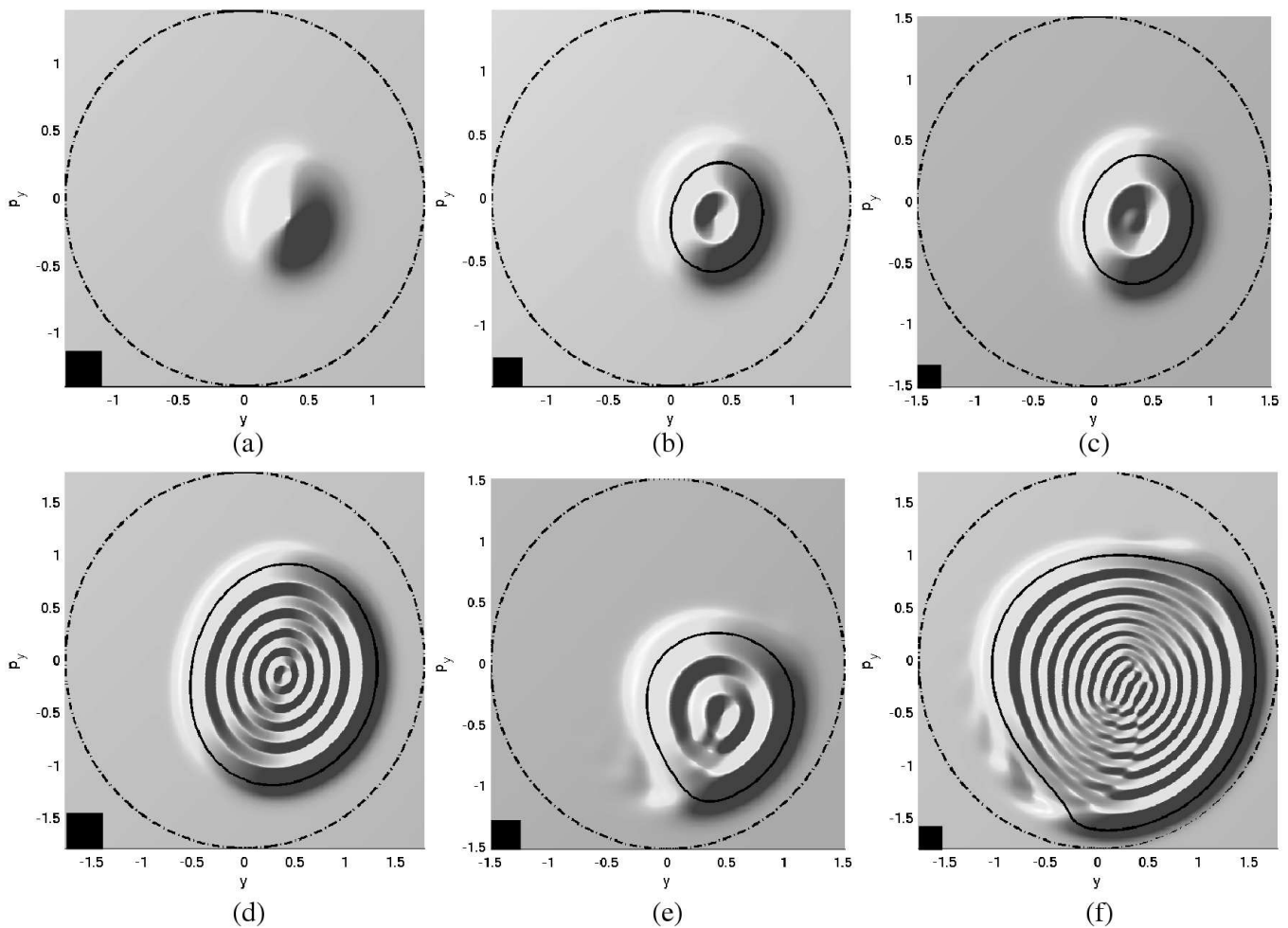}
\caption{\label{fig1} The Weyl symbol $\Wt_{\R}(\zetab,E)$
is shown for the potential in (\ref{toypot}). Cases (a) to (d)
have $\Omega=1$ and $\lambda=\mu=1/2$ and energies (a) $E=1.0$,
(b) $E=1.10$, (c) $E=1.15$ and (d) $E=1.6$. Cases (e) and (f)
have  $\Omega=1$, $\lambda=-1/2$ and $\mu=1$ and energies
$E=1.15$ and $E=1.6$ respectively.
Also shown are, the boundary of the allowed region
as a dashed curve and, in cases (b) to (f) where the energy is
above threshold, the boundary of the classically reacting region.
In each case a box of area $\hbar$ is shown on the bottom left.}
\end{figure*}

We write the scattering matrix in block form
\[
  S(E)=\left(
  \begin{matrix}
	r_{RR} & t_{RP}\\
	t_{PR} & r_{PP}
  \end{matrix}
  \right),
\]
where the subscripts $R$ and $P$ refer to reactant and product
channels respectively. For example, the transmission matrix $t_{PR}$
maps asymptotically incoming states on the reactant side to
asymptotically outgoing states on the product side. The theory we
describe is for the matrix
\[
  \hat{\mathcal{R}}(E)=t_{PR}^\dagger t_{PR}
  \label{eq:reaction}
\]
rather than for the scattering matrix itself. The matrix $\R(E)$ has an obvious
physical role determining state-specific reaction rates.
The transmission probability
for an incoming state labelled $\ket{\psi_n}$ on the reactant side
can be written as a matrix element
\begin{equation}
  p_n = \braopket{\psi_n}{\R(E)}{\psi_n}
  \label{eq:probabilityR}
\end{equation}
of $\R(E)$.
In going from the scattering matrix to $\R(E)$ we
lose information about phase and about the distribution of product
states but, as described in \cite{sC04}, uniform
semiclassical approximations for $\R(E)$ are expected to be considerably
simpler than those for $S(E)$.

Formally, we can think of $\R(E)$ as an operator acting on the Hilbert
space $\H_R^\in(E)$ of asymptotically propagating states in the incoming
reactant channel --- in this context we refer to it as the
{\it reaction operator} in the following. We
consider systems for which there are a finite number
$M$ of such states so $\H_R^\in(E)$ is finite-dimensional and $\R(E)$ can be
represented by an $M\times M$ matrix. The Wigner-Weyl correspondence
offers an alternative representation as a function on phase space
and the theory we outline in the coming sections is stated in those
terms. In the remainder of this section we describe how the connection is made
formally between the matrix representation and the representation
in  Wigner-Weyl correspondence, or the {\it Weyl symbol} of $\R(E)$.
In preparation for that discussion, let us first describe the phase space
on which these representations are defined.

As described in \cite{sc05} (see also \cite{Bog}), the natural classical
analog of the Hilbert space $\H_R^\in(E)$ is a {\it Poincar\'e section}
$\Sigma_R^\in(E)$ defined by fixing the energy $E$ and an asymptotically
large value of the reaction coordinate $x$. We use $(y,p_y)$ as
canonical coordinates on $\Sigma_R^\in(E)$ and we may alternatively
denote points in $\Sigma_R^\in(E)$ using the vector notation
\[
\zetab = \spinor{y}{p_y}.
\]
We define the {\it allowed region} of the $y$-$p_y$ plane by the condition
$p_y^2/2  + V_\infty(y) < E$ and note that, as usual in this sort
of correspondence, the dimension $M$ of
$\H_R^\in(E)$ is approximated by the Liouville area of this region
divided by $2\pi\hbar$.

To calculate the Weyl symbol $\Wt_{\R}(\zetab,E)$ of $\R(E)$,
we denote its individual matrix elements by
$
 R_{nm}(E) = \braopket{\psi_n}{\R(E)}{\psi_m}
$
and write
\begin{equation}\label{TWnm}
\Wt_{\R}(\zetab,E) = 2\pi\hbar \sum_{nm} R_{nm}(E) \Wt_{nm}(\zetab)
\end{equation}
where $\Wt_{nm}(\zetab)$ are the Weyl symbols of the projectors
$\ket{\psi_n}\bra{\psi_m}/2\pi\hbar$ (the factors of $2\pi\hbar$ are to
keep the notation consistent with standard practice for Wigner functions
in the case $n=m$). For the asymptotically harmonic potential in
(\ref{toypot}) the functions $\Wt_{nm}(\zetab)$ are known analytically
(see \cite{Ripamonti} for example).

The Weyl symbol $\Wt_{\R}(\zetab,E)$ of $\R(E)$ provides a remarkably
transparent means of visualising the quantum transmission problem
and of relating the scattering matrix to the geometry
of classical phase space. To illustrate, we show examples of
$\Wt_{\R}(\zetab,E)$ in Figure~\ref{fig1} for the model
potential in (\ref{toypot}) with energies at and above threshold.
These results have been obtained by first computing the scattering matrix
numerically using symplectic integration combined with the log
derivative method as described in \cite{dM86,dM95} and then using
(\ref{TWnm}). In each case we see that $\Wt_{\R}(\zetab,E)$
is effectively supported in a region of $\Sigma_R^\in(E)$,
which we can identify as the quantum-mechanically reacting
region. Indeed, by rewriting the reaction probability in
(\ref{eq:probabilityR}) (using standard properties of the Wigner-Weyl
correspondence) in the form
\begin{equation}
  p_n=\int\Wt_{\R}(\zetab,E)\Wt_{nn}(\zetab) \d\zetab,
\label{eq:aver}
\end{equation}
where $\d\zetab=\d y\d p_y$, and interpreting the Wigner function
$\Wt_{nn}(\zetab)$ as a phase space pseudodensity,
it is natural to identify $\Wt_{\R}(\zetab,E)$ as a probabability
of reaction as a function of phase space, albeit in an averaged
sense. Although the uncertainty principle prevents us from
defining a point-wise transmission probability in phase space,
we can construct linear combinations of incoming states with
a fixed total energy (as in (\ref{scatstate}) below) whose Wigner
function is supported within an area of $O(\hbar)$ in
$\Sigma_R^\in(E)$ and the appropriate modification of (\ref{eq:aver})
then gives the reaction probability as an average of
$\Wt_\R(\zetab,E)$ over that support.

For energies at or below threshold, transmission is controlled by
tunnelling and $\Wt_{\R}(\zetab,E)$ is supported in a phase
space region of area $O(\hbar)$, centred around an initial condition
that leads to an optimal tunnelling route. An explicit semiclassical
expression for $\Wt_{\R}(\zetab,E)$ in this case will be given later
and one can see for the threshold case in Figure~\ref{fig1}(a) that
$\Wt_{\R}(\zetab,E)$ is indeed peaked around a single point
in $\Sigma_R^\in(E)$.
As the energy increases above threshold, a classically reacting
region appears, initially centred on the orbit associated with optimal
tunnelling. The boundaries of the classically reacting regions
are indicated in Figures~\ref{fig1}(b) to (f) by continuous
closed curves (as the energy falls to the threshold case $E=1$,
the classically reacting region shrinks to a point corresponding to
the optimal tunnelling route and around which
 $\Wt_{\R}(\zetab,E)$ is concentrated). One can see in each case
that the reacting region closely matches the support of
$\Wt_{\R}(\zetab,E)$.

Before describing how $\Wt_{\R}(\zetab,E)$ is approximated semiclassically,
we should outline how the classically reacting regions in Figure \ref{fig1}
are defined.
The boundary of the classically reacting region in full phase space
is the stable manifold on the reacting side of a PODS in two dimensions
or more generally a NHIM if higher-dimensional problems are treated.
Extended into the incoming reactant channel, this stable manifold
defines a tube, the interior of which consists of classically reacting
trajectories and whose annular exterior in the classically allowed
region consists of trajectories which eventually return along the
outgoing reactant channel. A representation of this reacting region
in a Poincar\'e section $\Sigma_R^\in$ is obtained simply by taking
the a section of the tube of reacting trajectories at fixed energy and a fixed,
asymptotically large value of the reaction coordinate $x$. Since the tube
continues to evolve asymptotically (according to the dynamics of a
decoupled potential $V_\infty(y)$), the shape of a reacting region
defined in this way will depend on the value chosen for the coordinate
$x$. In models where the dynamics of the reacting region
is nonlinear --- or potentially even chaotic in problems of higher
dimension --- the shape of the reacting region will not have a limit
and becomes
ever more complicated as $x$ is brought to infinity. To obtain a fixed
asymptotic limit we therefore renormalise the dynamics by using
the decoupled evolution of the limiting potential $V_\infty(y)$
to map the asymptotic section back to one corresponding to a fixed
finite value of $x$. In making semiclassical comparisons the value
of $x$ used to define this final section is dictated by the conventions
used for the scattering matrix.

In the present case the asymptotic states in terms of which the
scattering matrix is defined are of the form,
\begin{equation}\label{scatstate}
\Psi_{n,E}(x,y) \sim \frac{\e^{ik_n x}}{\sqrt{\hbar k_n}}\psi_n(y),
\end{equation}
where $\psi_n(y)$ are the eigenfunctions of the decoupled potential
$V_\infty(y)$. The  phases of these scattering states are zeroed at
$x=0$ and asymptotic incoming states can be constructed by starting with
the transverse modes at $x=0$ and propagating them
backwards into the asymptotic region of the incoming channel using
dynamics defined by $V_\infty(y)$. To make a comparison
with the classical picture, the renormalisation of the classical
dynamics should therefore take an asymptotic Poincar\'e section back
to one defined by $x=0$. This is the convention used in Figure~\ref{fig1}
to compare ${\Wt}_{\R}(\zetab,E)$ with the classically reacting region.
Alternative phase conventions would lead to a reaction operator $\R(E)$
obtained by conjugation of the one we define by a unitary matrix
which is diagonal in the basis $\ket{\psi_n}$. This conjugation makes
no difference to the diagonal matrix elements $\braopket{\psi_n}{\R}{\psi_n}$
but is important for the appearance of the Weyl symbol in $\Sigma_R^\in$.
Choosing values other than $x=0$ for the reference section $\Sigma_R^\in$
would, for example, lead to a deformation of the Weyl symbol by the
asymptotically decoupled dynamics.

Note that this renormalisation procedure is simply means of
interpreting a term in the phase function in the classical
$S$-matrix \cite{MillerSmatrix} that fixes its asymptotic value.
By accounting for this term using a conjugation of the asymptotic
dynamics by the mapping in decoupled dynamics back to $x=0$, we can
incorporate everything about the classical $S$-matrix into a single
Poincar\'e mapping and present results in a more compact form,
as described more fully in the coming sections.

\section{Harmonic Approximation}
\label{harmonic}
In this section we describe a semiclassical approximation for $\R(E)$
based on dynamics linearised around an optimal tunnelling orbit.
Although less accurate and valid over a smaller range of energies
than the fully nonlinear theory described in the next section,
this harmonic approximation captures the essential
qualitative behaviour of $\R(E)$ and works well in the especially
interesting range of energies around threshold where classical
reaction switches on. It is also considerably simpler to apply
and can be expressed in closed form using easily obtained classical
data. Note that the term ``harmonic'' here refers simply to the fact
that linearised dynamics about a tunnelling orbit are used and has
nothing to do with the harmonic asymptotic behaviour of the model
we use for numerical illustration.

\subsection{Operator version}

\begin{figure*}
\includegraphics[width=18cm]{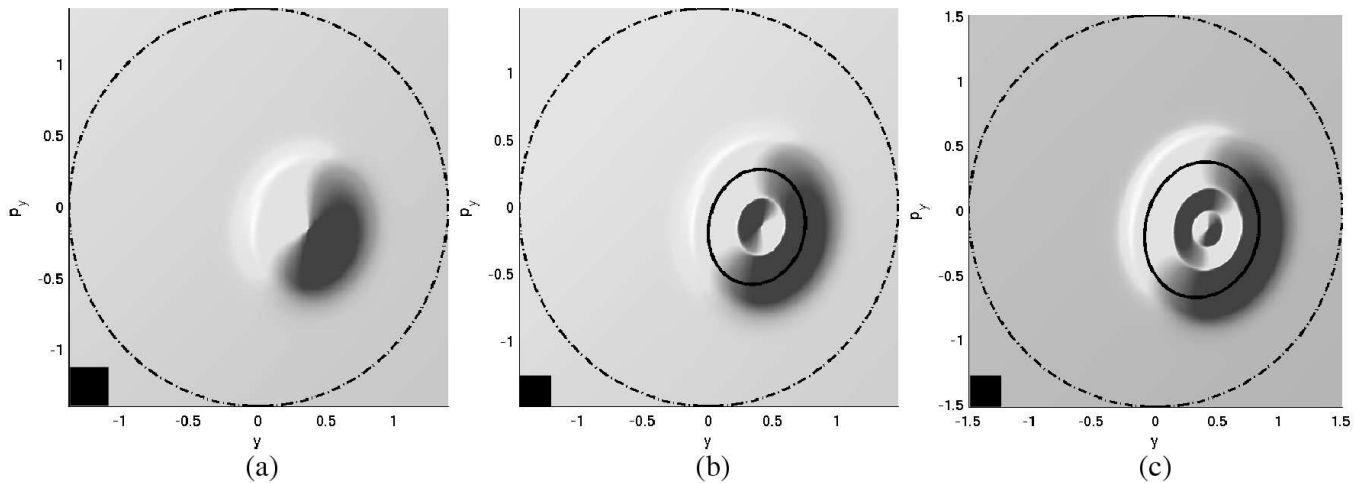}
\caption{\label{fig4}The harmonic approximation to $\Wt_\R(\zetab,E)$
is shown for energies (a) $E=1$, (b) $E=1.10$ and (c) $E=1.15$,
corresponding to the top row in Figure~\ref{fig1}}
\end{figure*}

A full description of the harmonic approximation to $\R(E)$ has been
given in \cite{sC04} and we refer there for details and a derivation
of the approach.
Here we simply summarise the important points. The operator $\R(E)$
is approximated by a formula
\begin{equation}
  \hat{\mathcal{R}}(E)\approx\frac{\hat{\mathcal{T}(E)}
  }{1+\hat{\mathcal{T}(E)}}
\label{eq:t2}
\end{equation}
where $\T(E)$ is a ``tunnelling operator'' constructed
from classical data. The elements needed to compute $\T(E)$ are as
follows.
\begin{itemize}
\item A complex periodic orbit $\gamma_E(t)$,
the ``instanton'', is found which encircles
the transition state in imaginary time. This orbit
can be found for energies above and below threshold
and its dynamical characteristics depend smoothly on energy there.
\item At the end of the previous section it was described
how  the boundary of the reacting region can be first extended
arbitrarily far into the asymptotic region and then renormalised
by mapping back to a section  $\Sigma_R^\in$ at $x=0$ using
decoupled dynamics, so that the shape remains fixed as
dynamics are extended into the asymptotic region. An analogous
renormalisation, illustrated in Figure~\ref{fig1},
is applied to $\gamma_E(t)$ so that an asymptotically fixed initial
condtion for it is defined in $\Sigma_R^\in$. Above threshold this
initial condition is near the centre of the reacting region.
\item The imaginary action of $\gamma_E(t)$ is denoted
$iK_0=\oint_{\gamma_E} \p\cdot\d\q$. We also denote
$\theta=K_0/\hbar$. Note that for energies above threshold
we have  $K_0<0$ while $K_0>0$ below threshold.
\item Linearised dynamics around $\gamma_E(t)$ are characterised
by a complex monodromy matrix $W$, which is routinely determined
as part of a numerical search for the orbit $\gamma_E(t)$. The eigenvalues
of $W$ come in real reciprocal pairs $(\Lambda,\Lambda^{-1})$,
which we order so that  $\Lambda>1$.
\item The matrix $W$ can be generated by using an elliptic
quadratic Hamiltonian $h(\zetab)$ for an imaginary time $-i\tau_0$.
Note that this Hamiltonian generates renormalised dynamics in the section
$\Sigma_R^\in$ and is therefore {\it not} simply a truncation
of the full Hamiltonian in the transition state region.
\item Canonical coordinates $(Q,P)$ are defined on $\Sigma_R^\in$ so
that
\[
h(\zetab) = \frac{\alpha}{2}\left(Q^2+P^2\right)
\]
and we have $\Lambda=\e^{\alpha\tau_0}$.
\end{itemize}
The quantum analog of the classical generating Hamiltonian  $h(\zetab)$
is denoted by $\hh$.

We can now write the tunnelling operator $\T(E)$ in the form
\begin{equation}\label{Tdef}
\T = \e^{-\theta-\tau_0\hh/\hbar},
\end{equation}
which, except for the prefactor $\e^{-\theta}$,
is an imaginary-time evolution operator generated by
$\hh$. Since $\hh$ is harmonic we can explicitly construct
its eigenstates $\ket{\varphi_k}$, with $k=0,1,\cdots$ and the
corresponding eigensolutions of $\T$ are
\[
\T\ket{\varphi_k}=\tau_k\ket{\varphi_k}
\]
where the eigenvalues
\begin{equation}
  \tau_k = \e^{-\theta}\Lambda^{-(k+\frac{1}{2})}
  \label{eq:deftau}
\end{equation}
are deduced simply by exponentiating the eigenvalues
of $\hh$.

\subsection{Weyl symbol}

It is shown in \cite{sC04} how closed form approximations
can be deduced for phase space representations of $\R(E)$ as a result
of substituting this exponentiated form for $\T(E)$ in  (\ref{eq:t2})
and resumming the geometric series $\R=\T-\T^2+\T^3-\cdots$. This leads
to an integral representation
\begin{equation}\label{intrep}
\R(E) = \frac{1}{2i}\int_C
\frac{\e^{-\rho(\theta+\tau_0\hh/\hbar)}}{\sin\pi\rho}\d\rho
\end{equation}
for $\R(E)$, in which the contour $C$ ascends just to the right
of the imaginary axis. Standard asymptotic approaches to this integral,
such as the method of steepest descent, allow explicit asymptotic
approximations to be written for $\R$ in various representations,
including for the Weyl symbol.

These expressions are especially useful to understand the detailed structure
of $\R(E)$ in phase space, but for the purposes of computing
$\R(E)$ for the parameter regimes we consider here, it suffices
to use an an eigenexpansion
\begin{equation}
  \hat{\mathcal{R}}(E)\approx\sum_{k}  r_k \ket{\varphi_k}\bra{\varphi_k}
  \label{eq:eigenexpansion}
\end{equation}
where
\begin{equation}
  r_k=\frac{\tau_k}{1+\tau_k}.
  \label{eq:defrk}
\end{equation}
The Weyl symbol of $\R(E)$ can, for example, be written as
\[
\Wt_{\R}(\zetab,E) \approx 2\pi\hbar \sum_k r_k(E) \tilde{\Wt}_{kk}(\zetab)
\]
where $\tilde{\Wt}_{kk}(\zetab)$ are the Wigner functions of the states
$\ket{\varphi_k}$ (and given analytically in \cite{Ripamonti} for example).
The tilde distinguishes these Wigner functions from those of the
basis states $\ket{\psi_n}$ of the scattering operator, which are
different.

The canonical coordinates $(Q,P)$ are centred on the
initial condition for $\gamma_E(t)$ in $\Sigma_R^\in$ and are such that
for energies just above threshold, the classically reacting region is
circular in the $(Q,P)$ plane. In the original coordinate system
$(y,p_y)$ these Wigner functions are translated, squeezed and rotated
so that their level curves are aligned with the approximately
elliptical reacting region. The resulting approximation for
$\Wt_{\R}(\zetab,E)$ therefore describes an elliptically-shaped
representation of the true quantum transmission problem

\begin{figure}
\includegraphics[width=8.5cm]{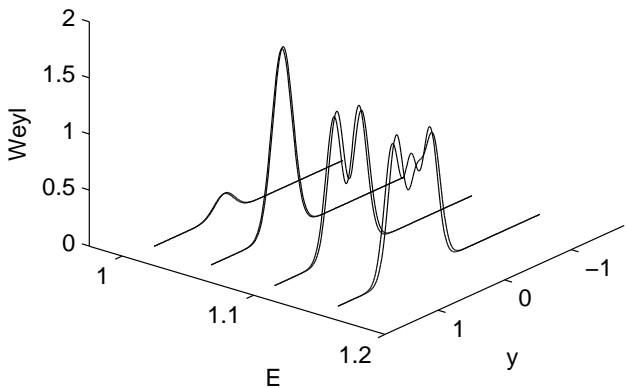}
\caption{\label{fig5} The harmonic approximation is compared with
exact results for $\Wt_\R(\zetab,E)$ at energies at and
just above threshold. These are obtained by sampling $\Wt_\R(\zetab,E)$
along a horizontal line through the centre of the reacting region
in $\Sigma_R^\in$ and plotting the result as a function of $y$.
Parameters in the potential are as in Figures~\ref{fig1}(a)-(d) and
\ref{fig4}. }
\end{figure}

The harmonically approximated Weyl symbol $\Wt_{\R}(\zetab,E)$ is
illustrated in  Figure~\ref{fig4} for three energies at and just
above threshold in the model potential with $\Omega=1$ and
$\lambda=1/2=\mu$. For comparison,  illustrations of corresponding
exact calculations can be found in the top row of  Figure~\ref{fig1}.
In general we find
that the harmonic approximation is in good quantitative agreement
with exact results at and below threshold. The threshold case in
Figure~\ref{fig4}(a), for example, is indistinguishable from the
corresponding exact result in Figure~\ref{fig1}(a) at the level
of graphical resolution used. As energy increases, the agreement
deteriorates so that noticable differences are visible
when $E=1.15$ (harmonic approximation in Figure~\ref{fig4}(c)
and exact calculation in  Figure~\ref{fig1}(c)). It should be emphasised,
however, that even then, the harmonic approximation captures the essential
qualitative features of  $\Wt_{\R}(\zetab,E)$.

In order to make a closer comparison between exact and harmonic
results, we show one-dimensional sections through the Weyl symbol in
Figure~\ref{fig5}.
In each case  $\Wt_{\R}(\zetab,E)$ is sampled along a horizontal
line through the centre of the reacting region in $\Sigma_R^\in$
and plotted as a function of the $y$ coordinate. There is good
quantitative agreement in cases (a) and (b) where the enrgy is at and
just above threshold. At higher energies the harmonic approximation
captures the support of the quantum-mechanically reacting region well
but details of the Wigner function do not match at the centre
of the reacting region. It should be remarked, however, that oscillations
in the Weyl symbol are sensitive to nonlocal changes in phase space
and discrepencies at the centre of the reacting region may not
have a strong effect on averaged reaction probabilities as expressed in
(\ref{eq:aver}).

\begin{figure}
\includegraphics[width=8.6cm]{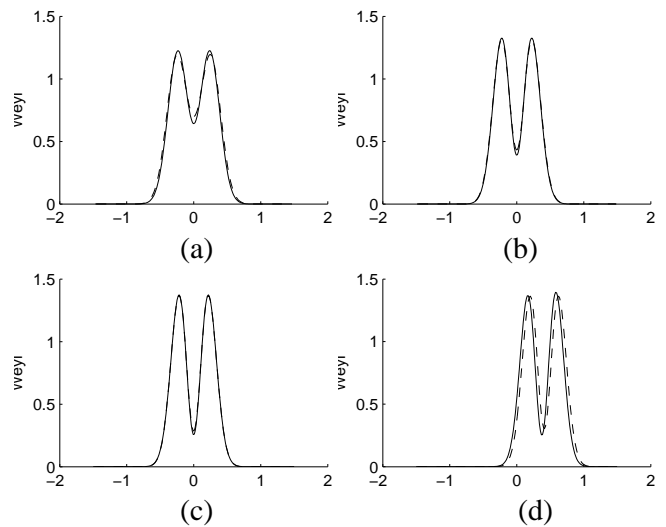}
\caption{\label{fig6}The harmonic and exact
Weyl symbols are compared for different potentials with $\Omega=1$ and
(a) $\lambda=-\ha$, $\mu=0$; (b) $\lambda=0$, $\mu=0$; (c)
$\lambda=\ha$, $\mu=0$; (d) $\lambda=\ha$, $\mu=\ha$.
In each case the energy is chosen so that the classically reactive
flux has the fixed value $N_{\rm cl}(E)=2$.}
\end{figure}

Similar one-dimensional sections are shown in Figure~\ref{fig6}
which illustrate the harmonic approximation for different parameter
sets. In each of these the energy is chosen so that the cumulative
reactive flux
\[
N_{\rm cl}(E) =\frac{1}{2\pi\hbar}\oint_{\rm PODS}\p\cdot\d\q
=\frac{E-1}{\hbar\sqrt{\Omega^2+\lambda}}
\]
is fixed (at the value 2). Figures~\ref{fig6}(a), (b)
and (c) show cases where $\mu=0$ and the potential has a symmetry
in $y$. Figure~\ref{fig6}(a) has a negative value of $\lambda=-1/2$
for which an adiabatic approximation assuming fast transverse
dynamics in the barrier region would not be expected
to apply. Figure~\ref{fig6}(b) is the separable case $\lambda=0$
and in Figure~\ref{fig6}(c) we have $\lambda=1/2$. Note that separability
does not confer a particular computational advantage in this approach,
nor does it lead to particularly better accuracy of the approximation.
In Figure~\ref{fig6}(d), an example is shown in which $\mu=1/2$ and
the potential is neither separable nor symmetric in $y$. The approximation
works less well in that case. This is not unexpected because corrections
to the harmonic approximation will be quartic rather than cubic
in a symmetric problem but we note that there is still good agreement.

\section{Non-Linear Approximation}\label{nonlinear}
Although the harmonic approximation captures the essential
qualitative features of quantum transmission  and works well
quantitatively near threshold, we can achieve greater range of
applicability and significantly improved numerical agreement if
we use fully nonlinear dynamics around the orbit $\gamma_E(t)$.
The price to be paid for this improvement is that
the resulting calculation is significantly more involved.
The greatest impediment is that, although the tunnelling operator defined
by nonlinear evolution can be routinely approximated semiclassically,
we do not know at present how to write semiclassical approximations
for the operator $(1+\T)^{-1}$ directly in terms of classical
orbits. In this paper we simply
use numerical inversion of the matrix representation of $1+\T$.
Before describing this procedure, it is helpful to
describe how nonlinear calculation is incorporated in
the operator $\T$. This is done in section~\ref{primsec} below, followed
by a description of the uniform calculation in  section~\ref{unisec}.

\subsection{Primitive approximation}\label{primsec}
At energies below threshold the imaginary action of the orbit
$\gamma_E(t)$ is positive, that is $\theta>0$, and the exponential
prefactor $\e^{-\theta}$ in (\ref{Tdef}) makes $\T$ small. We may therefore
approximate the reaction operator directly by the tunnelling
operator, giving
\[
\R(E) \approx \T(E),
\]
which we refer to as the {\it primitive approximation}. The primitive
approximation is easily extended beyond the immediate neighbourhood of
$\gamma_E(t)$. Instead of letting $\T$ be the evolution operator corresponding
to the classically linear evolution defined by $W$, as we did in the
previous section, we let it be the quantum version of a nonlinear map
in $\Sigma_R^\in$.

Initial conditions near $\gamma_E(t)$ in $\Sigma_R^\in$ can be followed
over a sequence of time evolutions similar to those of  $\gamma_E(t)$
itself until they return to $\Sigma_R^\in$, defining a surface-of-section
mapping which we denote by
\[
\F:\Sigma_R^\in\to\Sigma_R^\in.
\]
As with conventional return maps, $\F$ defines a canonical
transformation on $\Sigma_R^\in$, except that it is {\it complex},
in general taking real initial conditions to complex images. Despite this
complexity, the evolution has a quantum analog as an evolution operator,
and this is the tunnelling operator $\T$.

\begin{figure}
  \begin{center}
	\includegraphics[width=7.5cm]{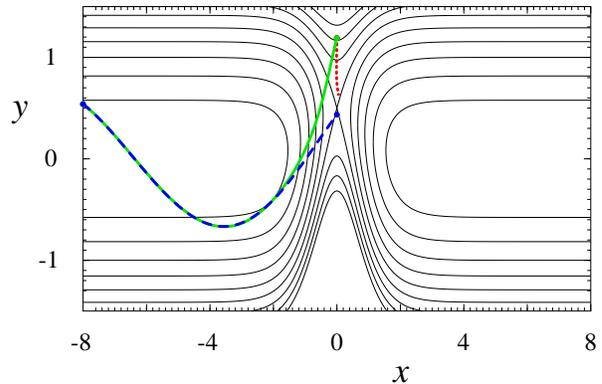}
  \end{center}
  \caption{\label{fig:mp} The projection onto real configuration space
of a typical midpoint orbit is shown. This orbit can be obtained by
continuously deforming the initial conditions for the orbit shown in
Figure~\ref{fig:orbit}. The dotted segment corresponds
to evolution over the imaginary part of the time contour and the dashed
segment corresponds to renormalisation by the uncoupled dynamics. The
symmetry $\zetab_B=\zetab_A^*$ means that the real part of the orbit
shown here is self-retracing, with the orbit turning back on itself
at the end of the dotted segment.}
\end{figure}

With suitable modifications to take account of the complexity of the mapping,
standard semiclassical approximations that are applied to evolution operators,
such as the Van Vleck formula, can be used to approximate $\T$. Here we
focus on an approximation derived in \cite{mB89} for the Weyl symbol
of an operator (see also \cite{sC99}). For the Weyl symbol of the
operator $\T$ we write
\begin{equation}
\Wt_{\T}(\zetab,E)\approx\frac{\e^{-\mathcal{A}(\bm{\zeta},E)/\hbar}}
{\sqrt{\det(W_{AB}+I)/2}},
  \label{defWeyl2}
\end{equation}
where $\A$ and $W_{AB}$ are calculated from a {\it midpoint orbit}
$\zetab_A\to\zetab_B$ which is defined by the conditions
\begin{eqnarray}\label{defmp}
\zetab &= &\ha\left(\zetab_A+\zetab_B\right)\nonumber\\
\zetab_B &=& \F(\zetab_A).
\end{eqnarray}
That is, the point $\zetab$ at which  the Weyl symbol is to be
evaluated is the midpoint of $\zetab_A$  and $\zetab_B$,
where  $\zetab_A$ evolves into  $\zetab_B$ under the return map. The
exponent $\A(\zetab,E)$ is such that
\[
i\A(\zetab,E) = \int_{\zetab_A}^{\zetab_B}\p\cdot\d\q - p_y(y_B-y_A)
\]
and the matrix $W_{AB}$ is a linearisation the map $\F$ around the
orbit $\zetab_A\to\zetab_B$.

It can be shown \cite{sc05} that the complex conjugate of
the map $\F$ is its inverse, $\F^*=\F^{-1}$, and from this
a number of important symmetries follow which guarantee
that $\Wt_{\T}(\zetab,E)$ is a real-valued function on
$\Sigma_R^\in$ that is peaked around the initial condition
for $\gamma_E(t)$, which we denote by $\zetab_0$ in the following.
On a formal level, the real-valuedness of
 $\Wt_{\T}(\zetab,E)$ follows from the observation that $\T$
is Hermitian which, as discussed in \cite{sC99}, is a quantum analog
of the property $\F^*=\F^{-1}$. It is instructive, however, to see
how the real-valuedness of the semiclassical approximation to
$\Wt_{\T}(\zetab,E)$ follows directly from the symmetries of the
midpoint orbit.

 First we note that, given a midpoint orbit
$\zetab_A\to\zetab_B$ for a real-valued $\zetab$, then
$\zetab_B^*\to\zetab_A^*$ is also a midpoint orbit (for the
same $\zetab$). This can be seen by conjugating the relations in
(\ref{defmp}) and using
$\zetab_B^*=[(\F(\zetab_A)]^*=\F^*(\zetab_A^*)=\F^{-1}(\zetab_A^*)$
to deduce that $\zetab_A^*=\F(\zetab_B^*)$. It turns out in fact that
these two midpoint orbits coincide, so
\[
\zetab_B = \zetab_A^*
\]
and
\[
\Re\,\zetab_B = \zetab = \Re\,\zetab_A.
\]
This is easily confirmed for the linearised map (replacing $\F$
by multiplication by $W$ and using $W^*=W^{-1}$) and therefore holds
for the nonlinear map if $\zetab$ is close enough to $\zetab_0$.
The condition $\zetab_B = \zetab_A^*$ can only be violated if
a bifurcation is encountered and a more detailed analysis shows
that this corresponds to the condition $\det(W_{AB}+I)=0$,
which would lead to a caustic in (\ref{defWeyl2}). We will
assume in this paper that no such caustics are encountered in
the region of $\Sigma_R^\in$ which dominates reaction.

An example of a full trajectory corresponding to a typical
midpoint orbit is illustrated in Figure~\ref{fig:mp}. Because the
initial conditions are complex, the coordinates of the trajectory are
generically complex over its length, even along the segments
which have been obtained by deformation of the real segments
of $\gamma_E(t)$. A consequence of the symmetry $\zetab_B = \zetab_A^*$,
however, is that the time contours can be chosen so that the
second half of the trajectory reverses the complex conjugate
of the first half. A projection onto real configuration space,
for example, is self retracing. We find as a result that the action
is purely imaginary and the exponent $\A(\zetab,E)$ is a positive
real number. We also find that $W_{AB}^*=W_{AB}^{-1}$ and because
$\det W_{AB}=1$ this means that the amplitude term
$\det(W_{AB}+I)$ in (\ref{defWeyl2}) is real (and positive). Therefore
$\Wt_{\T}(\zetab,E)$ is a positive real-valued function with a maximum
at $\zetab_0$ (for which we have $\A(\zetab_0,E)=K_0(E)$).

\begin{figure}
\includegraphics[width=7cm]{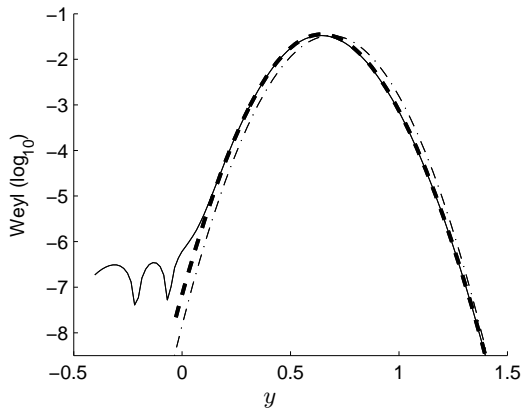}
\put(-100,-7){$y$}
\caption{\label{fig7} For an energy $E=0.985$ just below threshold,
a comparison is given on a logarithmic scale between the  exact Weyl
symbol (continuous curve), the nonlinear primitive approximation of
Equation~(\ref{defWeyl2}) (heavy dashed curve) and the harmonic
primitive approximation of Equation~(\ref{primharm}) (light
dashed curve). The parameters here are $\Omega=1$, $\lambda=-1/2$,
and $\mu=1$. The  nonlinear primitive result works well over
a large region of phase space but does not capture the oscillations
in the tail of the exact calculation, which may well be
a signature of nonintegrable complex-dynamical
effects of the type described in \cite{aY00,Takahashi,TYK}}
\end{figure}

The harmonic
approximation can be recovered by expanding the exponent to second
order about $\zetab_0$ and approximating $W_{AB}$ by $W$,
giving \cite{sC04}
\begin{equation}\label{primharm}
\Wt_\T(\zetab,E)\approx
\frac{\e^{-\theta-[(2/\beta)\tanh\beta/2] \tau_0 h(\zetabs)/\hbar}}
{\cosh\beta/2}.
\end{equation}
A comparison is given in Figure~\ref{fig7} between this approximation,
the fully nonlinear approximation of (\ref{defWeyl2}) and exact results.
Although the harmonic approximation works well near the maximum of the
Weyl symbol, the fully nonlinear result works better over a larger range.
We note however that there is qualitative deviation even from the nonlinear
approximation in the deep tunnelling regime where the exact
calculation shows significant oscillations not captured by the
harmonic or nonlinear approximations. Similar oscillatory structure, or
``fringed tunnelling''  in the scattering matrix  has been explained
in \cite{aY00,Takahashi,TYK} on the basis of nonintegrable complex
dynamics and has
been shown to involve mechanisms that also show up in chaotic tunnelling.
It seems likely that  the oscillations in Figure~\ref{fig7} have a
similar origin but we have not preformed a detailed analysis.
It will be an interesting problem in the future to combine the
inherently nonintegrable mechanism in \cite{aY00,Takahashi,TYK} with the
uniform approximations illustrated here.

\subsection{Uniform approximation}\label{unisec}
Although we now have an explicit closed-form semiclassical approximation
for $\T$, no equivalent result is currently available for
the uniformisation $\T/(1+\T)$ because inversion of the operator
$1+\T$ cannot be done simply.
 In this paper we simply adopt a hybrid approach
which combines semiclassical approximation of $\T$ with numerical inversion
of $1+\T$. Although not a fully semiclassical
method, this will allow us to verify that (\ref{eq:t2}) gives an accurate
reproduction  of quantum transmission.

We first represent $\T$ as a matrix in the same asymptotic basis
$\ket{\psi_n}$ as used for the scattering matrix. We denote individual
matrix elements by $T_{nm}=\braopket{\psi_n}{\T}{\psi_m}$ and compute
them using the Wigner-Weyl calculus by writing
\begin{equation}
  T_{nm}(E)= \int \Wt_{nm}(\zetab) \Wt_{\T}(\zetab,E)\d\zetab
  \label{eq:tkl2}
\end{equation}
and approximating $\Wt_{\T}(\zetab,E)$ using (\ref{defWeyl2}).
We emphasise that while $\T$ is almost diagonal in the basis
$\ket{\varphi_k}$ of eigenstates of the generating Hamiltonian $\hh$,
the same is not true in the basis $\ket{\psi_n}$ unless the potential is
separable.
The integral is performed numerically and the resulting $M\times M$
matrix for $\T/(1+\T)$, whose elements are denoted $R_{nm}^{\rm sc}(E)$,
is also computed numerically. This integration is not difficult since a grid
on $\Sigma_R^\in$ is easily filled by using Newton integration to
step the midpoint orbit, starting with the known solution corresponding
to $\gamma_E(t)$ at $\zetab_0$ (for which $\zetab_A=\zetab_0=\zetab_B$).
Once the elements $R_{nm}^{\rm sc}(E)$ are known,
the Weyl symbol for $\T/(1+\T)$ is obtained by replacing
$R_{nm}(E)$ with  $R_{nm}^{\rm sc}(E)$ in (\ref{TWnm}).

\begin{figure}
\includegraphics[width=8.5cm]{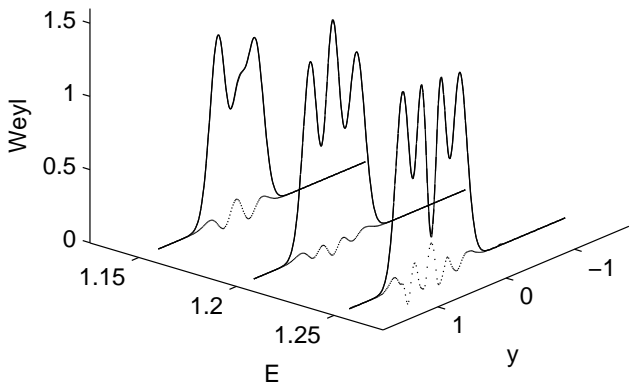}
\caption{\label{figconj1} The uniform nonlinear approximation is compared with
the exact Weyl symbol for a series of energies and the same parameters
as used in Figure~\ref{fig5}. It is difficult the distinguish
the exact and approximate results at the level of graphical resolution
used, but in each case the difference multiplied by $10$ is shown
underneath as a dotted curve.}
\end{figure}

We find excellent agreement between the semiclassically computed
Weyl symbol $\Wt_{\R}(\zetab,E)$ and the exact result. A nonlinear
version of Figure~\ref{fig4} is indistinguishable from the exact results
shown in the top row of Figure~\ref{fig1} and therefore not shown.
Instead we compare  in Figure~\ref{figconj1} horizontal slices of the
Weyl symbol through the reacting region, in the same manner as  in
Figure~\ref{fig5}. The potential used is the same
as in Figure~\ref{fig5} and the energies treated extend somewhat higher
above the threshold. The exact and semiclassical results cannot be
distinguished at the resolution used, so the difference scaled by
a factor of $10$ is also shown. We find similarly good agreement for
other parameter sets we ahve investigated and we note that the quality
of the approximation
does not require special features of the classical dynamics such as symmetry,
separability or adiabatic separation of transverse from reaction
degrees of freedom.

We should remark that the current hybrid implementation of the
nonlinear calculation is cumbersome and is harder to apply
further above the barrier where the larger region of integration
demands that we extend the midpoint trajectory deeper
into complex phase space. We have not, for example, reproduced
the results on the second row of Fig.~\ref{fig1} using this method.
The purpose of this calculation is to show that the nonlinear
uniform result derived  in \cite{sc05} provides an accurate
description of $\R(E)$ in the model considered and that the method
therefore deserves further exploration. Even though
numerical inversion was used in applying the formalism, it is built
entirely on a semiclassical approximation for the tunnelling
operator $\T(E)$ and we expect that any subsequent fully semiclassical
implementation will be equally accurate.

We also remark that the current hybrid method is theoretically
clumsy and obscures somewhat the deeper connections between the
quantum results and the underlying classical geometry. For example,
it would be especially interesting to characterise the behaviour
of $\Wt_\R(\zetab,E)$ at the boundary of the classically reacting region
where trajectories approach the PODS (or NHIM in higher dimensions)
along its stable manifold and where the classical reaction probability
drops sharply from $1$  to $0$. Such an analytical approximation was found in
\cite{sC04} for the harmonic version in which $\Wt_\R(\zetab,E)$
is approximated as an integral of the Airy function near the boundary
of classical reaction. Investigation is currently underway into a
method to derive similar results in the nonlinear case on the basis of
generating $\T$ as in equation (\ref{Tdef}), but with an
anharmonic generator $\hh$ computed using classical normal form
theory. Ultimately it should be possible to describe explicitly how
the quantum reaction probability varies across the boundary of the
classically reacting region in terms of trajectories which approach
the complexified PODS along its stable manifold and evolve along
it before returning to the asymptotic Poincar\'e section.

\section{Conclusions}
We have successfully treated quantum transmission across a
multidimensional barrier using uniform semiclassical
approximation. The method applies generically around a
threshold energy and does not rely on specific features of the
classical dynamics such as separability or the existence
of action angle variables. In its fully nonlinear
incarnation the method gives an accurate description of reaction
probabilities in phase space and makes a striking connection
between the quantum scattering problem and the geometry of classical
reaction. We expect that it will
work equally well in higher-dimensional problems, even in cases
where the incoming states are chaotic in the transverse dynamics.

Although we have shown that the fully nonlinear version works
well, in doing so we have resorted to numerical methods which
are not in the spirit of semiclassical approximation. The theory
therefore needs further development in order to achieve a fully
semiclassical description of the emerging reacting region.
One promising approach which is currently under investigation
is to use classical normal form theory to generate dynamics
around the orbit $\gamma_E(t)$ using a nonlinear extension
of the generator $h(\zetab)$. Many of the explicit analytical
approaches used in the harmonic case might then be adapted to
the nonlinear approximation. In particular this is expected
to produce a detailed analytical description of the Weyl symbol
at the boundary of the reacting region which calls on intrinsic
geometrical features of (the stable manifold of) the NHIM.

A second aspect of the calculation which deserves further
attention is the treatment of rotational degrees of freedom
in fully three-dimensional models of chemical reaction.
Although at one level this is simply a question of applying
the results here individually to symmetry-reduced phase spaces
for given angular momentum quantum numbers, there are interesting
and nontrivial problems in describing the quantum-classical
correspondence compactly in operator form. This is an especially
interesting issue for reactions which proceed through a
collinear mechanism since the collinear configurations
are a singular part of the classical reduction process.

\vspace{60pt}

\noindent{\large\sc Acknowledgements}\\
\noindent CSD is supported by an EPSRC studentship and SCC acknowledges
support by the European Network MASIE, Contract No. HPRN-CT-2000-00113..
\indent

\end{document}